\documentclass[doublecol,
article,
groupedaddress,
nofootinbib,
 amsmath,amssymb,
 aps,
prb,
]{epl2}

\usepackage{graphicx}
\usepackage[utf8]{inputenc}
\usepackage{dcolumn}
\usepackage{bm}

\title{Detection of vortex coherent structures in superfluid turbulence}

\author{E. Rusaouen$^1$, B. Rousset$^2$ and P.-E. Roche$^1$}
\institute{%
\inst{1} Institut NEEL, CNRS, Universit\'e Grenoble Alpes, F-38042 Grenoble, France\\
\inst{2} SBT/INAC CEA, Universit\'e Grenoble Alpes, F-38054 Grenoble, France\\
}%

\date{\today}


\abstract{
Filamentary regions of high vorticity irregularly form and disappear in the turbulent flows of classical fluids. We report an experimental comparative study of these so-called ``coherent structures'' in a classical versus quantum fluid, using liquid helium with a superfluid fraction varied from 0\% up to 83\%. The low pressure core of the vorticity filaments is detected by pressure probes located on the sidewall of a 78-cm-diameter Von K\'arm\' an cell driven up to record turbulent intensity ($R_\lambda \sim \sqrt {Re} \simeq 10000$ ). The statistics of occurrence, magnitude and relative distribution of the filaments in a classical fluid are found indistinguishable from their superfluid counterpart, namely the bundles of quantized vortex lines. This suggest that the internal structure of vortex filaments, as well as their dissipative properties have a negligible impact on their macroscopic dynamics, such as lifetime and intermittent properties.}

\begin{document}
\maketitle


\section{Introduction}

\subsection{Motivation}

Turbulent flows of water, air or other classical fluids are populated by so-called ``coherent structures''. These structures are  localized in space and characterized by an organized flow motion. In particular, worm-shaped regions of high vorticity -often referred as ``vortex filaments''- irregularly spring up, and after a life-time significantly larger than their turn-over time,  destabilize and vanishe \cite{siggia1981numerical,she1990intermittant,vincent1991satial,DouadyFilament:PRL1991,jimenez1998characteristics}.

A few numerical studies of superfluid helium have shown that bundles of quantum vortex lines should be the counterparts of classical vortex filaments in quantum fluids. The formation of such bundles in a freely evolving quantum fluid have been recently reported  in Ref. \cite{Baggaley_Coherentvortexstructures_EPL2012}. This  result was preceded by a number of numerical studies where an external field was promoting  the formation of vortex bundles in a superfluid  (e.g. see Ref. \cite{Kivotides:2006,Morris:PRL2008}).

The motivation of the present study is to detect experimentally coherent structures in quantum turbulence.

\subsection{Experimental context}

The comparison between classical and quantum (or superfluid) turbulence has focused a lot of attention over the last years \cite{BarenghiSkrbekSreenivasan_IntroPNAS2014}. Regarding experimental studies of turbulent fluctuations, the situation is contrasted \cite{spectra:PNAS2014}.
On the one hand, several similarities have been reported 
including on velocity spectra\cite{Maurer1998,Salort:PoF2010} and energy transfer between eddies of different sizes \cite{Salort:EPL2012}. 
On the other hand, differences between classical and quantum turbulences are reported when vorticity (instead than velocity) is directly or indirectly probed, by spectral measurements of the vortex line density \cite{RocheVortexSpectrum:EPL2007,Bradley:PRL2008} and by visualization of reconnections of individual vortices \cite{Bewley:Nature2006,Paoletti:2008p488}. 

In this context, coherent vortex structures are interesting objects to compare classical and quantum turbulence. Indeed, a bundle of quantum vortices is an intermediate structure living between the quantum scales (where a quantized vortex line can move without dissipation) and macroscopic scales (where classical turbulent properties are expected).

\subsection{Methodology}

We use liquid helium $^4$He, both above its superfluid transition (where it is a classical fluid) and below it, where it acquires properties of a quantum fluid \cite{VanSciverLivre2012,DonnellyLivreVortices}.
In the later case, according to the two-fluid model of Landau and Tisza, it behaves as an intimate mixture of a "normal" fluid and a "superfluid", which are coupled  by a mutual friction force. The normal fluid follows the Navier-Stokes equation, while the superfluid has zero viscosity and can be described as a tangle of quantized vortex lines. In the zero temperature limit, the normal fluid density (volumetric mass) $\rho_n$ vanishes and $^4$He becomes a pure superfluid. Conversely, near the transition temperature ($\simeq 2\,K$), the superfluid density $\rho_s = \rho - \rho_n$ vanishes. In the present study, the superfluid fraction $\rho_s / \rho$  varied from 0\% to 83\% ($2.46\,K \ge T \ge \,1.58 K$).

To detect coherent vortex structures, we look for the low pressure appearing in their core due to centrifugal force. This pressure depletion can be assessed from the Poisson equation for pressure $p$ in an incompressible flow \cite{bradshaw1981note}, derived by taking the divergence of Navier-Stokes equation (a generalization for compressible flow is proposed in \cite{HorneComment1992}):
\begin{equation}
\Delta p =  \frac{\rho}{2} \left( \omega^2 - \sigma^2 \right)
\label{eq:Poisson} 
\end{equation}
\noindent where $\rho$ are the fluid density, $\omega$,  and $\sigma$ are the flow vorticity and rate of strain defined as
\begin{equation}
 \omega^2 = \frac{1}{2} \sum _{i,j} \left( \partial_i v_j - \partial_j v_i   \right)^2
\end{equation}
\begin{equation}
 \sigma^2 = \frac{1}{2} \sum _{i,j} \left( \partial_i v_j + \partial_j v_i   \right)^2
\end{equation}
By analogy with electrostatics, equation \ref{eq:Poisson} shows that a localized region of high vorticity is a (negative) source term for pressure\footnote{contrary to a frequent assumption, $ \omega^2$ and $\sigma^2$ don't balance each other on average in closed flows\cite{Raynal:PoF1996}.}.
The technique of tracking low pressure spikes to detect coherent structures has been widely used in classical turbulent flows, in particular the Von K\'arm\' an geometry 
 (eg. see Ref.~\cite{fauve1993pressure,cadot:pof1995characterization,roux1999detecting,chainais:PoF1999intermittency,TitonCadot:PRE2003,burnishev:pof2014}).
 In practice, a pressure transducer is imbedded in the sidewall of the cell; when a vortex filament passes by the probe, the resulting negative spike greatly exceeds in magnitude the standard deviation of the pressure fluctuations generated by the``background'' turbulence. Thus, the vortex filament can be detected.

Generalization of this  equation in a quantum fluid at finite temperature is straightforward in the framework of HVBK equations, discussed in \cite{Hills:1977p314}. In this approach, the superfluid tangle is coarse-grained into continuous velocity $\vec{v}_s$ and vorticity $\vec{\omega_s}$ fields. The detail of individual vortices is lost but the resulting equation for the superfluid  can account for fluid motion at scales much larger than the typical inter-vortex distance. The HVBK equations are an Euler equation  for the superfluid (underscript $s$) and a Navier-Stokes equation for normal fluid (underscript $n$), both coupled together :
 \begin{equation}
\rho_s \big[ (\partial \vec{v}_s/\partial t)+ (\vec{v}_s \cdot  \nabla) \vec{v}_s  \big] = 
- \frac{\rho_s}{\rho} \nabla p  + \rho_s S  \nabla T    -  \vec{F} 
\label{eq:euler} 
\end{equation}
 \begin{equation}
\rho_n \big[ (\partial \vec{v}_n/\partial t)+ (\vec{v}_n \cdot  \nabla) \vec{v}_n  \big] = 
- \frac{\rho_n}{\rho} \nabla p   - \rho_s S  \nabla T                    + \vec{F} + \mu \nabla^2 {\vec{v}_n}
\label{eq:ns} 
\end{equation}

\noindent where $\mu$ is  the dynamic viscosity, S is the entropy, and where the coupling term $\vec{F}$ accounts mutual coupling.

Assuming incompressibility, and taking the divergence of the sum of Eq. \ref{eq:euler} and \ref{eq:ns}, one gets a generalized Poisson equation in the two-fluid model :

\begin{equation}
\Delta p =  \frac{\rho_s}{2} \left( \omega_s^2 - \sigma_s^2 \right) + \frac{\rho_n}{2} \left( \omega_n^2 - \sigma_n^2 \right)
\label{eq:Poisson2} 
\end{equation}

\noindent The above equations shows that negative pressure spikes in a quantum fluid remain markers of high vorticity regions. Superfluid and normal fluid vorticities are probed simultaneouly, and weighted in proportion of the density of each fluid. Note that the low pressure on individual quantum vortices has been invoked to explain the trapping of light particles along vortices (see \cite{Bewley:Nature2006,Sergeev:2009p4065,Guo:PNAS2014} and references within).

\section{Experimental set-up}

\subsection{The Von K\'arm\' an flow}

\begin{figure}[ht]
\includegraphics[width=0.95\linewidth]{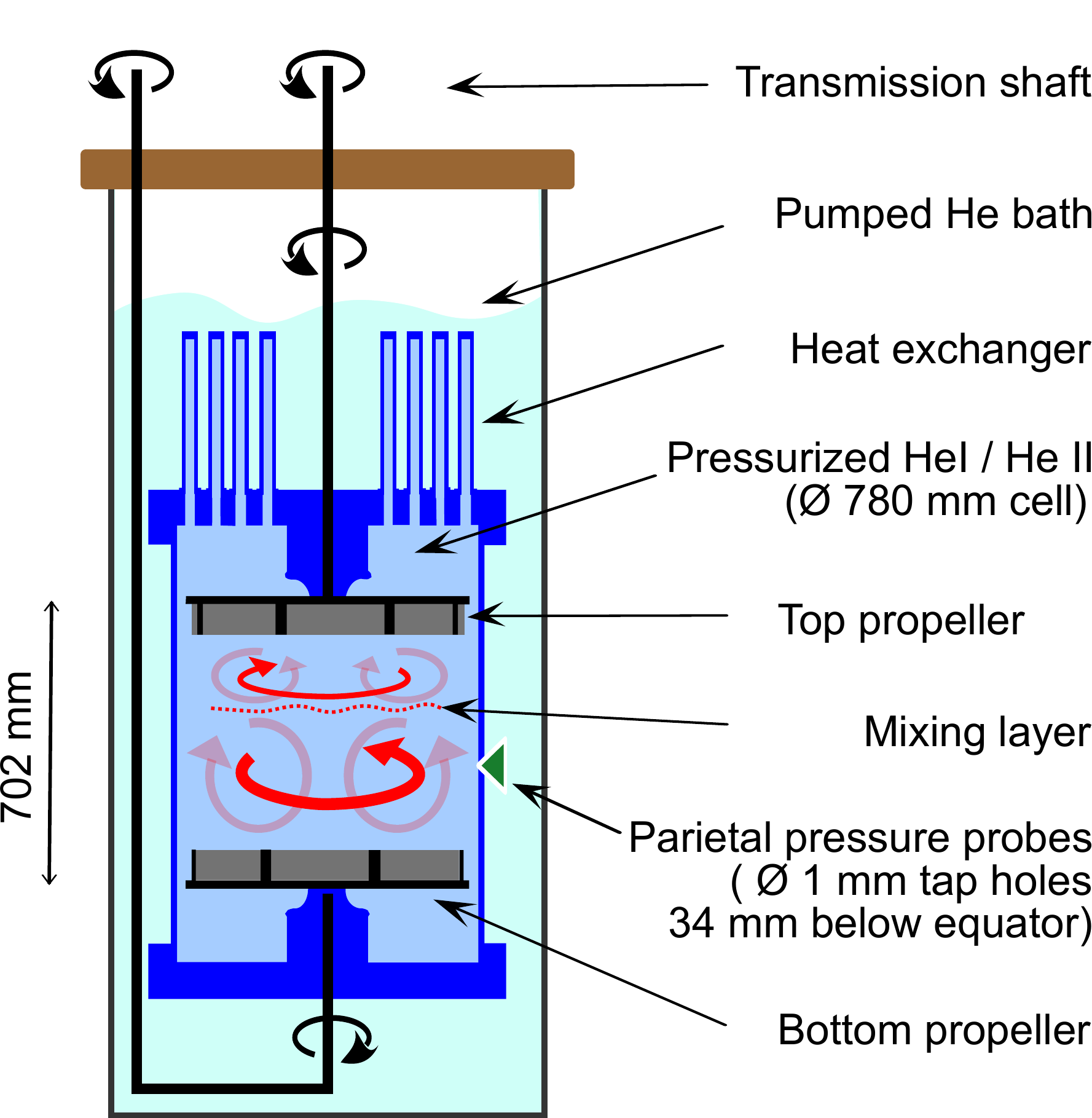}
\caption{\label{fig:schema} Schematic of the experiment.}
\end{figure}

The Von K\'arm\' an flow used for this experiment has been extensively described in a dedicated paper \cite{Rousset:RSI2014}. We only recall below its main specifications, see figure \ref{fig:schema}.

The liquid helium $^4$He used in this experiment was sequentially set to temperatures of 2.4 K,  2.1 K and 1.6 K, that is both above and below the superfluid transition temperature ($T_{\lambda} \simeq  2.15 K$ at 3 bars). These three temperatures correspond respectively to superfluid fractions of 0\%, 19\%  and $80\%$ at the pressures of interest (see Table~\ref{tab:table1}). The pressurization of the flow prevents occurrence of cavitation for all flow conditions.

The flow is enclosed in a 780-mm-diameter cylindrical vessel and it is mechanically stirred by two co-axial bladed-disks of radius $R=360 mm$, located $702 mm$ away,  counter-rotating in this work. The 8 blades on each disk are curved, and the direction of rotation is such that the convex side of the blades moves into the fluid. This specific direction is chosen because it results in a stable large scale circulation between the disks\cite{Rousset:RSI2014}.

Such a stirring gives rise to two counter-rotating subflows separated by a mixing layer, as depicted in Fig. \ref{fig:schema}.
The (mean) position of this mixing layer is determined by the relative angular velocities $\Omega_b$ and $\Omega_t$ of the bottom and top disks. For exact counter-rotation ($\Omega_b=\Omega_t$), the mixing layer is located at mid-height. In this study, we set  $\Omega_b > \Omega_t$, to position the mixing layer above the mid-plane away from the probes which are located 34~mm below this mid-plane. The relative angular velocity of the disks is characterized by  :
\begin{equation}
\theta = \frac{\Omega_b - \Omega_t}{\Omega_b + \Omega_t}
\end{equation}
The parameter $\theta$ was set to 11-12\% and 20\% to probe the flow at two distances from the mixing layer.
In classical Von K\'arm\' an flow, the Reynolds number is often defined as :

\begin{equation}
Re = \frac{\rho R^2 (\Omega_b + \Omega_t)}{2 \mu} = \frac{\rho R^2 \overline{\Omega} }{ \mu}
\end{equation}
\noindent where  $\rho$ is the density of the fluid and $\overline{\Omega} $ is the mean angular velocity.
For our purposes, this definition remains a convenient control parameter below the superfluid transition. Indeed, 
at large scales, the superfluid and normal fluid are strongly locked by the  mutual coupling force
which make them behave as a single fluid of viscosity $\mu$  \cite{Morris:PRL2008,Roche2fluidCascade:EPL2009}.

The flow parameters $\theta$ and $Re$ used in the present study are given in Table~\ref{tab:table1}. We stress that this study is performed at ultra large Reynolds number, of order $Re\simeq 10^8$ rarely reached in laboratory conditions. Following \cite{mordant1997characterization}, the typical Taylor microscale Reynolds number can be assessed from $Re$ as $R_\lambda \simeq \sqrt(Re) \simeq 10000$.

\begin{table*}
\caption{\label{tab:table1}%
Characteristics of the times series.
}
\vspace{0.5em}
\begin{tabular}{ccccccccc}  

\textrm{Superfluid }&
\textrm{Temperature}&
\textrm{Pressure}&
\textrm{Reynolds}&
\textrm{Rotation}&
\textrm{Mean}&
\textrm{Azimutal}&
\textrm{Skewness}&
\textrm{Flatness}\\

\textrm{fraction}&
\textrm{}&
\textrm{}&
\textrm{number}&
\textrm{dissym.}&
\textrm{rotation}&
\textrm{velocity}&
\textrm{}&
\textrm{}\\

\vspace{0.5em}

\textrm{$\rho_s / \rho$}&
\textrm{[K]}&
\textrm{[Bar]}&
\textrm{Re}&
\textrm{$\theta$}&
\textrm{$\overline{\Omega}$ [rad/s]}&
\textrm{$V^\star$ [m/s]}&
&\\

0 \%   & $2.42 \,\,\,\,(>T_\lambda)$& 3.4  & $ 5.5 \times 10^7$ & 0.20 &  8.2      & 1.9  & -0.64 & 5.3 \\
0 \%   & $2.41 \,\,\,\,(>T_\lambda)$& 3.4  & $ 5.5 \times 10^7$ & 0.12 &  8.2      & 1.3  & -1.59 & 14.4 \\
0 \%   & $2.46 \,\,\,\,(>T_\lambda)$& 3.6  & $ 6.6 \times 10^7$ & 0.12 &  10.2   & 1.3  & -1.62 & 12.9 \\
19 \% &$ 2.10 \,\,\,\,(<T_\lambda)$ & 2.7 & $ 5.9 \times 10^7$ & 0.20 &  5.7     & 1.4  & -0.48 & 4.6 \\
19 \% & $2.10 \,\,\,\,(<T_\lambda)$ & 2.7 & $ 8.6 \times 10^7$ & 0.12 &  8.3     & 1.2  & -1.50 & 15.6 \\
19 \% & $2.10 \,\,\,\,(<T_\lambda)$ & 2.8 & $ 1.1 \times 10^8$ & 0.12 &  10.2   & 1.3  & -1.63 & 14.0 \\
79 \% & $1.64 \,\,\,\,(<T_\lambda)$ & 3.0 & $ 1.3 \times 10^8$ & 0.20 &  8.3     & 1.9  & -0.45 & 4.5 \\
79 \% & $1.64 \,\,\,\,(<T_\lambda)$ & 3.0 & $ 1.3 \times 10^8$ & 0.11 &  8.2     & 1.1 & -1.55 & 13.3 \\
83 \% & $1.58 \,\,\,\,(<T_\lambda)$ & 3.1 & $ 8.9 \times 10^7$ & 0.11 &  5.7     & 0.8 & -1.98 & 17.8 \\

\end{tabular}
\end{table*}

\section{Instrumentation}

\subsection{The parietal pressure probes}

Fluctuations of parietal pressure are monitored at two locations, both 34~mm below the mid-plane and at 80~mm from each other (measured along the sidewall circumference). At each location, a differential transducer senses the pressure difference between an orifice in the sidewall and a pressure reference.

The pressure reference is low-pass-filtered by an hydraulic impedance so that its mirrors the static pressure inside the flow, and follows its possible slow drift. 
From the spectral analysis of the measured pressure fluctuations, this lower cut-off frequency of the probe is estimated to be significantly lower than 100 mHz.

The orifice in the flow sidewall is a square-edge 1-mm-diameter hole, perpendicular to the wall, with an effective depth of around 20~mm. The membrane of the pressure transducer is mounted at the end of this connecting pipe. 
The Helmholtz resonance is close to 1\,kHz and mechanical vibrations of the transducers are damped using a mechanical filter.

In practice, the largest useful frequencies of the measured signal was not limited by the probe itself but by broadband pressure oscillations in the flow, in the hundreds of Hertz range. Those oscillations were probably originating from the cryogenic system maintaining the experiment cold.

\subsection{Electronics and acquisition}

Each piezo-resistive pressure transducer consists in a Wheatstone bridge laying over a deflecting membrane. Each bridge is polarized by a battery-based $\sim 350$\,mA current source. The bridge output voltage  is amplified using a low noise instrumentation preamplifier (0.6 $nV/\sqrt{Hz}$, model EPC1-B). A  8th-order linear-phase anti-alias filter at frequency $f_c$ (Kemo 1208/20/41LP) is inserted before an 18-bits acquisition board (National Instrument 6289). Acquisitions are performed at sampling frequency 20 kHz (with $f_c=6$\,kHz) and last between 25 and 45 min, except for a few sampled at 1 kHz (with $f_c=200$\,Hz) for practical reasons. All times series are post-processed by a numerical low-pass filter at 160 Hz to avoid possible post-processing artifact caused by the Helmholtz resonance.

\section{Results}

\subsection{Detection of coherent structures in turbulent superfluid}

\begin{figure}[ht]
\includegraphics[width=0.95\linewidth]{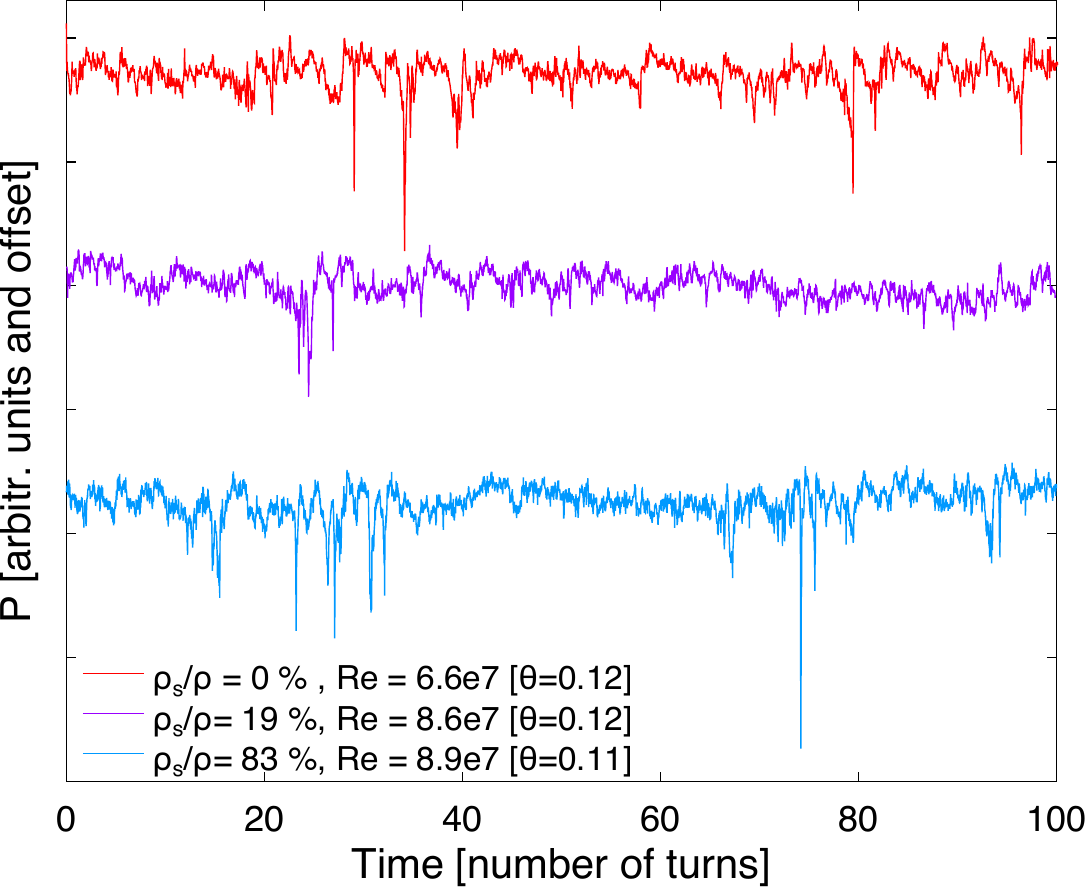}
\caption{\label{fig:timeSeries} Pressure time series at 3 temperatures for roughly similar forcing. The superfluid fraction ranges from 0\% to 84\%. Time on the x axis is rescaled by the mean rotation time $2\pi /  \overline{\Omega}$ of the disks. The sharp depressions are interpreted as the signature of vortical coherent structures passing over the pressure tap.} 
\end{figure}

We first discuss the classical flow regime ($\rho_s / \rho=0$).
The red time series plotted on Figure \ref{fig:timeSeries} illustrated the recording of several sharp depressions during 100 rotations of the disks. The time axis is scaled  by $2\pi /  \overline{\Omega}$ so that it corresponds to a number of turns of the disks.

Two possible artifacts of the measurements are acoustic noise within the fluid and mechanical noise propagating along the mechanical structure of the experiment. Pressure fluctuations were simultaneously recorded from two nearby sensors (as previously done in \cite{cadot:pof1995characterization}, for example), and were compared. Most depressions are only captured by one probe, which would not be the case if they were caused by an external noise source. Occasionally,  depressions are recorded by both probes with mean delays consistent with the mean direction of the flow, which confirms that the measured signal corresponds to localized coherent structures carried in the fluid.

Assuming a passive transport of the coherent structures between the two probes, the delay can be interpreted as a "time-of-flight" and gives the local flow (azimutal) velocity  $V^\star$ using the $8 cm$ probe separation. It is found in the $m.s^{-1}$ range, as given in Table ~\ref{tab:table1}. With $V^\star=1.6 m.s^{-1}$ and taking 160 Hz as the effective noise-free probe dynamics, we find a noise-free effective probe  resolution of $1 cm$ but  wavelet analysis of the raw time series (without the 160 Hz low-pass filter) allows to track the signature of the depression nearly up to the $\simeq 1 kHz$ probe resonance frequency, showing that the coherent structures can be at least as thin as $1.6 m.s^{-1} /1 kHz \simeq 2 mm$, to be compared with the large scale  $L$ of such Von K\'arm\' an flows  (\cite{burnishev:pof2014}).

\begin{equation}
L \simeq R/2 \simeq 200 mm,
\end{equation}
and to rough estimates of the Taylor and Kolmogorov dissipative scales $\lambda$ and $\eta$ based on the homogeneous isotropic turbulence equations.
\begin{equation}
\lambda \sim L \cdot \sqrt{10 / Re^\star } \simeq 0.2 mm
\end{equation}
\begin{equation}
\eta \sim L / {Re ^\star}^{3/4} \simeq 10^{-3} mm
\end{equation}
\noindent where we took $Re^\star = L V^\star  \rho / \mu \simeq 1.4 \cdot 10^{7}$. Surely, the flow is neither homogeneous nor isotropic, but these equations can still provide useful orders of magnitude, and show that the present probe is partly resolving the inertial range of the turbulent cascade, which extends from $\sim L$ down to $\sim 10 \eta$.

We now address the superfluid regime. Figure \ref{fig:timeSeries} illustrates two typical times series with superfluid fractions of $\rho_s / \rho  = 19 \%$ and $83\%$ acquired at Reynolds numbers similar to the classical regime ($Re=7.10^7 \pm 16\%$). As in the classical case, sharp depressions are found. No qualitative difference is found between the classical and superfluid regimes when all the acquired time series are scrutinized.

To the best of our knowledge, this is the first experimental evidence of coherent structures detected in a turbulent superfluid. 
We present below a quantitative analysis of the strength, density spatial distribution of those coherent structures with respect to their classical counterpart.

\subsection{Histogram of pressure : density and strength of coherent structures}

\begin{figure}[ht]
\includegraphics[width=0.95\linewidth]{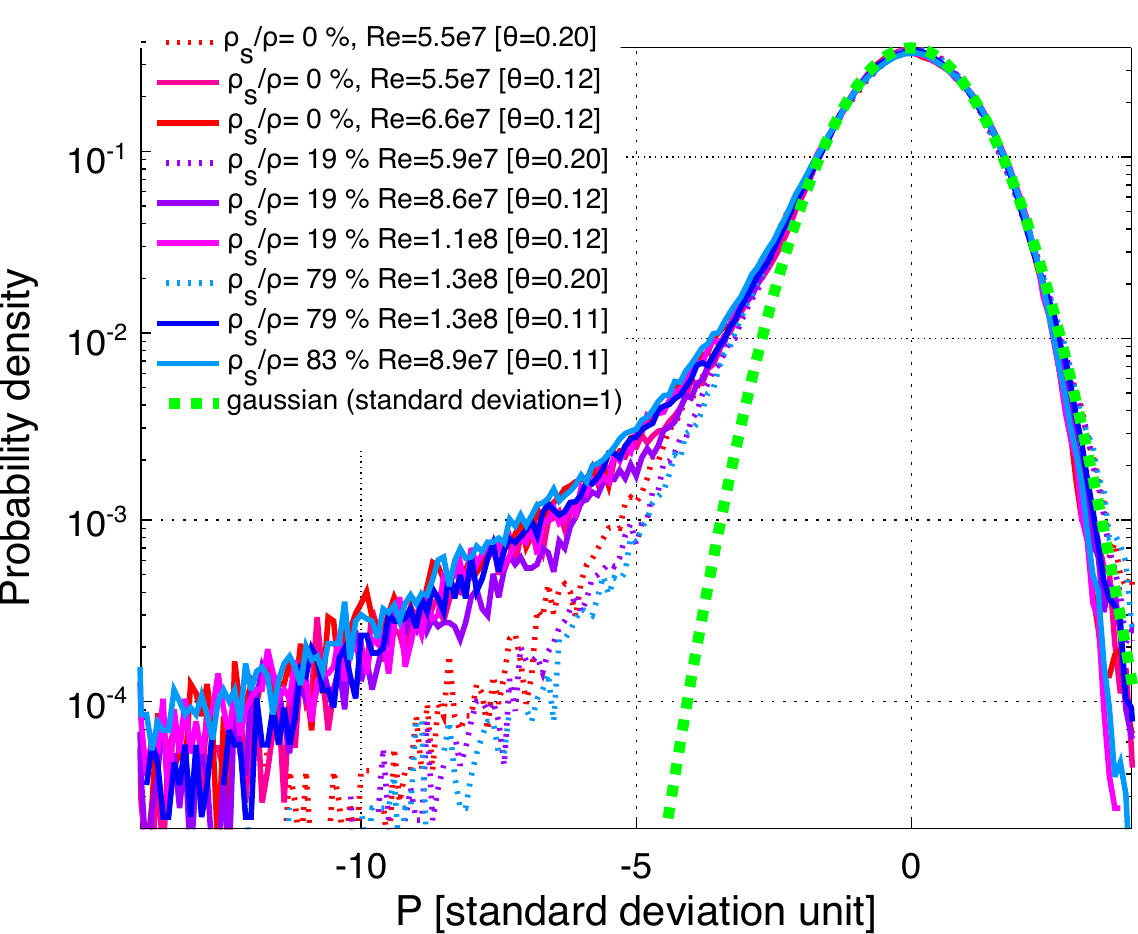}
\caption{\label{fig:hist} Probability density function (pdf) of the pressure fluctuations normalized to unity standard deviation.}
\end{figure}

Figure \ref{fig:hist} shows the probability density functions (pdf) of pressure time-series normalized by the standard deviation of their positive pressure fluctuations. 
The pdf shape is compatible with the description given in classical turbulence literature  for Von K\'arm\' an flows\cite{fauve1993pressure,cadot:pof1995characterization,chainais:PoF1999intermittency,TitonCadot:PRE2003,burnishev:pof2014}. It can be \textit{approximated} as gaussian complemented with a long exponential tail associated to the rare but intense negative pressures spikes associated with the coherent structures.
Such skewed pressure pdf have been reported in a number of classical turbulent flows, for instance in homogeneous isotropic turbulence \cite{metaisJFM:1992,pumir1994pressure}, along the centerline of pipes\cite{Lamballais:PRE1997} and in jets \cite{tsuji2003similarity} \footnote{in boundary layers more symmetrical pdf can be found, see e.g. \cite{Lamballais:PRE1997,Tsuji:2007p78}}.
One advantage of the Von K\'arm\' an geometry over these other flows is the efficient generation of vortex filaments in its mixing layer, and the resulting significant enhancement of the pressure skewness compared to the background skewness resulting from the quadratic velocity dependence of pressure\cite{holzer1993skewed}.

Whatever the superfluid fraction and Reynolds number, all the pdf corresponding to a given  $\theta$ are found to collapse, up to our statistical uncertainty. In other words, the density and strength of coherent structures are found independent of the superfluid fraction from 0\% up to 83\% of superfluid. This is the second important result of this study. When $\theta$  is lowered, the mixing layer gets closer to the probes and the density of coherent structures increases. It suggests that the mixing layer is an intense source of coherent structures, both in classical and superfluid turbulence. The dependence with the distance to the mixing layer can then be understood as the result of the finite life time\cite{TitonCadot:PRE2003} of the vortical coherent structures. This provides an indirect indications that the lifetime of the coherent structures is similar in the classical and superfluid cases.

The asymmetry and flatness of the pdf can be assessed quantitatively from two statistical quantities : the skewness and kurtosis of the pressure fluctuations. They are respectively defined as the centered third and fourth moment of the fluctuations normalized by their standard deviation.

\begin{figure}[ht]
\includegraphics[width=0.85\linewidth]{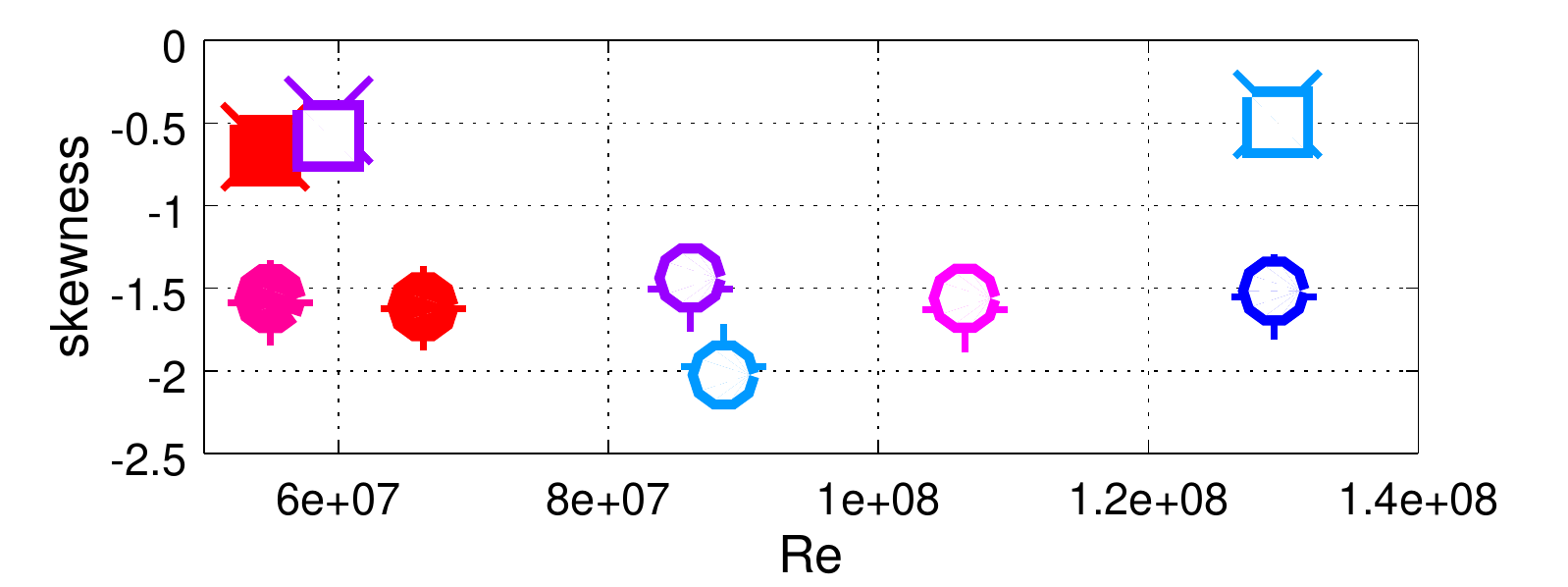}
\includegraphics[width=0.85\linewidth]{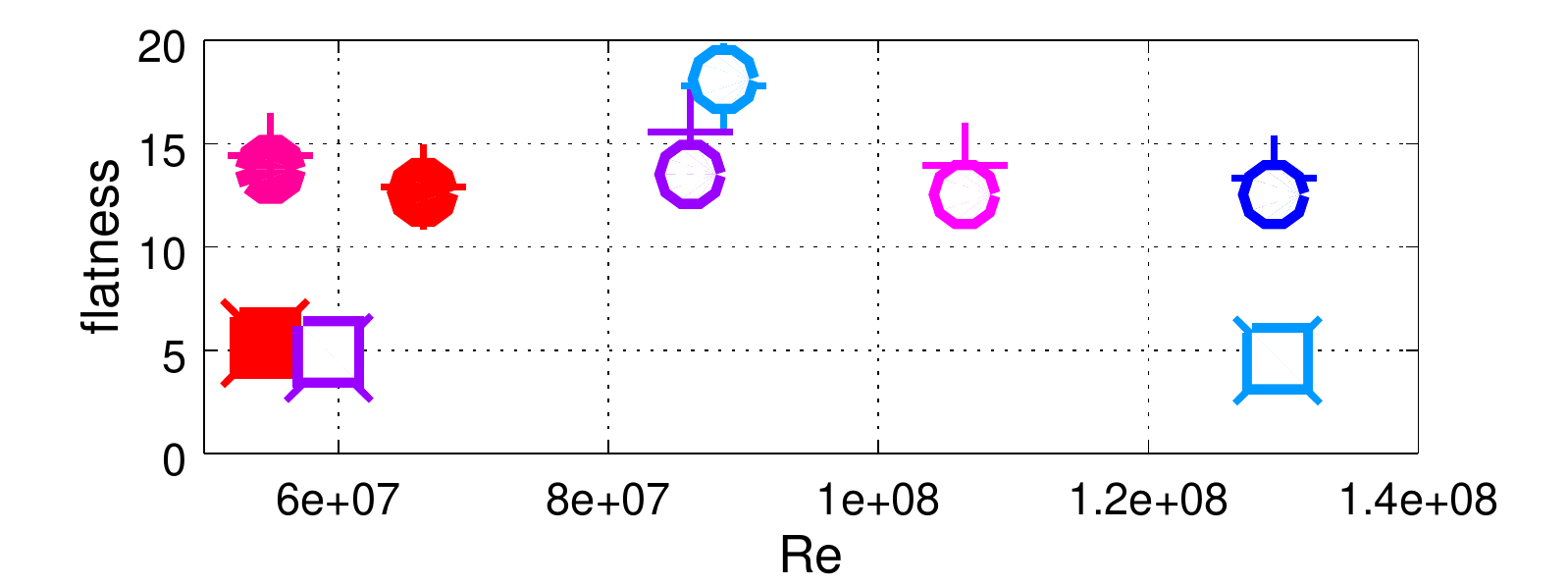}
\caption{\label{fig:skewness} Upper (lower) plot : skewness (flatness) of pressure fluctuations. The open (full) symbols correspond to measurements in superfluid (in classical liquid helium). The square-shaped (circle-shape) symbols are for a differential rotation parameter of $\theta = 0.2$ ($\theta = 0.11-0.12$). The crosses and pluses symbols correspond to the probe-bandwidth check with 53Hz low-pass filtering (see text).}
\end{figure}

Figure \ref{fig:skewness} shows the measured skewness and flatness (kurtosis) below and above the superfluid transition temperature. The numerical values are given in  Table~\ref{tab:table1}. Application of an additional $160\,Hz / 3 \simeq 53\,Hz$ low-pass filter on the time series don't alter significantly those quantities suggesting that we don't have time resolution issues. For a given value of $\theta$, no Reynolds number dependence emerges from our measurements when $Re\sim 10^8$ is varied by a factor 2.3, justifying a-posteriori that the definition of a  Reynolds number below the superfluid transition is not critical in the present study. 
On the contrary, the dependence of both parameters versus $\theta$ is around a factor 3.

\subsection{Spatial distribution of superfluid coherent structures}

\begin{figure}[ht]
\includegraphics[width=0.95\linewidth]{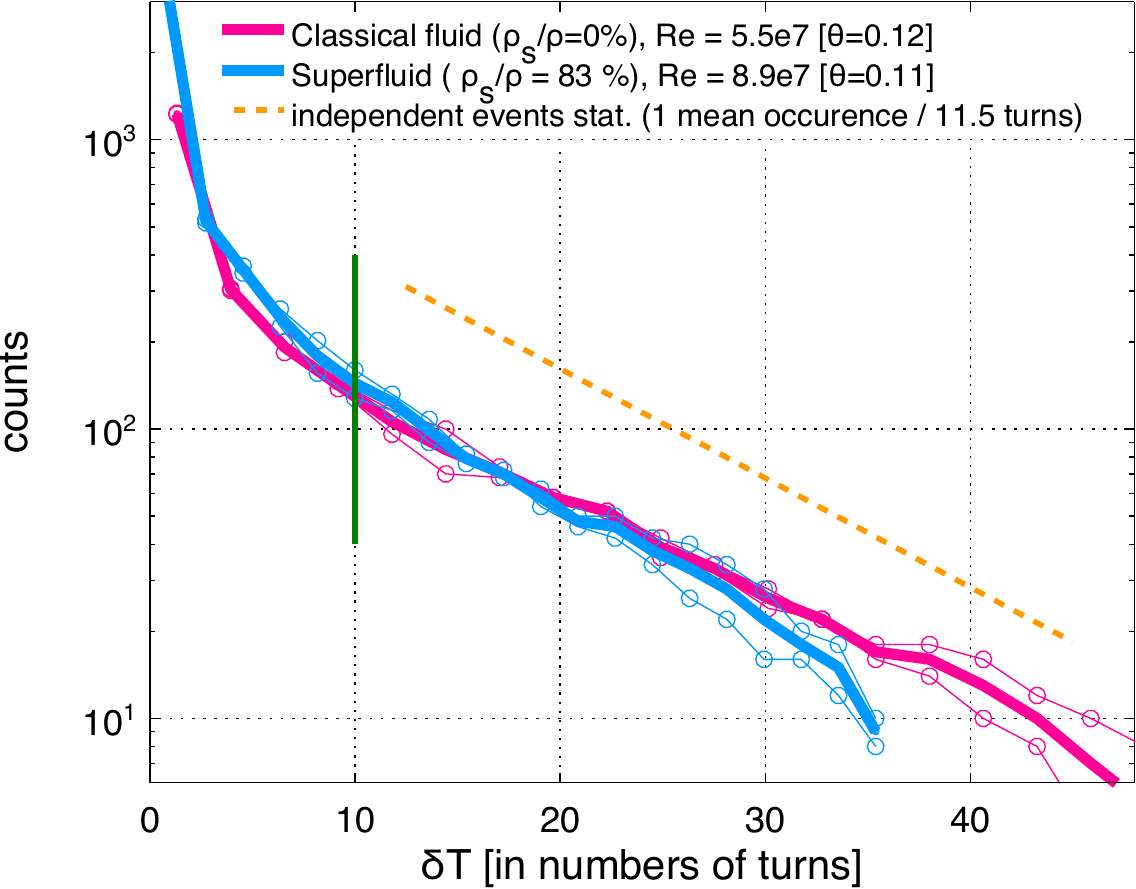}
\caption{\label{fig:ecart} Histogram of the intervals between successive coherent structures which are larger that $\delta T$. To improve statistical convergence, the statistics from two pressure taps (thin lines) have been averaged (thicker line). The dash line corresponds to the expected dependence of independent events with a mean separation of 11.5 mean rotations (see text).
}
\end{figure}

To go one step further in the comparison of coherent structures, we now address their relative spatial distribution in the classical and superfluid regimes. To this end, we focus on the statistics of time interval $\delta T$ between two consecutive coherent structures passing by one probe. We need to chose an arbitrary criterion for identification of coherent structures. Several criteria have been proposed and studied in the classical turbulence literature, with little incidence in the respective conclusions (eg. see \cite{cadot:pof1995characterization,roux1999detecting,abry1994analysis,chainais:PoF1999intermittency}). Following \cite{chainais:PoF1999intermittency}, we choose a pressure threshold at -3 in standard deviation units. Larger thresholds of 4 and 5 were also tested and gave compatible results but with a worst statistical convergence. In Figure \ref{fig:ecart}, the y-axis represents the number of intervals between successive coherent structures which are larger than $\delta T$ (x-axis). For best convergence, the longest time series at temperatures corresponding to 0 \% and 83 \% superfluid have been chosen and the times series from the two probes (thin lines) were averaged together (thick lines).

If coherent structures were fully independent from each other, we expect a Poisson statistics for the intervals $p(\delta T) \sim e^{-\delta T/\tau}$. By integration, the probability of interval larger than $\delta T$ is proportional to  $\tau e^{-\delta T/\tau}$. This exponential law accounts reasonably well for the results for intervals $\delta T$ longer than a characteristics correlation time of $\sim10$ mean rotation periods, in good agreement with classical turbulence literature\cite{abry1994analysis,chainais:PoF1999intermittency}. A fit gives a mean separation time  $\tau = 11.5 \pm 1.5$ in units of rotation period. For shorter intervals, the statistics is no-longer exponential. This reveals a trend for coherent structures to cluster, which is found similar in the classical and superfluid cases. 
In a frozen turbulence picture, this result means that the spatial distribution of the coherent structures is found similar in classical and quantum flows.

\section{Concluding remarks}

If the pressure probes were able to resolve individual quantum vortices, dissipative scales or the genuine pressure profile of a vortex bundle, some differences between measurements in a classical and in a quantum flows would be apparent.  Obviously, the resolution of present probes is not such, but we showed that it is sufficient to clearly detect the individual coherent structures, from their measured (low-pass-filtered) pressure profile. Thus, the statistics of occurrence and strength of coherent structures could be characterized and we found that they are statistically indistinguishable when measured in a classical flow and with a superfluid fraction of 19\% and 79\% to 83\%. In other words, the microscopic differences in internal structures of classical vorticity filament and superfluid vortex bundles do not prevent both types of coherent structures to recover similar macroscopic properties. 

Among the perspectives, it would be interesting to relate this findings to 
the unexpected $f^{-5/3}$ vortex line spectra \cite{RocheVortexSpectrum:EPL2007}, which have been interpreted as passive scalar spectra postulating that a large amount for vorticity was localized at small scales and carried by the flow \cite{RocheInterpretation:EPL2008,Salort:EPL2011}. The presence of vortex bundles could support very much this interpretation (for an alternative interpretation, see \cite{Nemirovskii:PRB2012}).
Another interesting perspective is to explore temperatures around 1.9K where a singular behavior has been numerically predicted for intermittency \cite{BoueIntermittency:PRL2013,ShuklaPandit:PRE2016}, but not yet evidenced experimentally\cite{Maurer1998,SalortIntermi:ETC13_2011}.
A third perspective would be understand the dissipative interaction between the bundles of superfluid vortices and the (possibly overlapping) filaments of normal fluid.

\section{Acknowledgements}

Financial support from EC Euhit project (WP21) is acknowledged, and special thanks its coordinator E. Bodenschatz for his initiative. We also thank members of the SHREK collaboration, with whom the facility was designed\cite{Rousset:RSI2014}, to M. Bon Mardion for facility operation, to A. Girard for Euhit aspects,  to P. Diribarne and M. Gibert for support in data-logging flow parameters and to B. Hébral for discussions and proof reading. We warmly acknowledge the help from O. Cadot in understanding better the origins of the skewness of pressure, and the feed-back from Y. Tsuji.

\bibliographystyle{eplbib}

\end{document}